\documentclass{appolb}
\usepackage{epsfig}

\begin{document}
\eqsec  
\title{ ROTATION CURVES OF SPIRAL GALAXIES: INFLUENCE OF MAGNETIC FIELDS AND 
ENERGY FLOWS 
\thanks{This research is partially supported by the Polish State Committee 
for Scientific Research (KBN), grants 2P03B 110 24 and PBZ-KBN-054/P03/02}%
}
\author{  M. Kutschera$^{a,b}$ and J. Jalocha$^a$
\address{
$~^a$Institute of Physics, Jagiellonian University\\
Reymonta 4, 30-059 Krak\'ow, Poland\\
$~^b$H.Niewodnicza\'nski Institute of Nuclear Physics\\
Radzikowskiego 152, 31-342 Krak\'ow, Poland}
}
\maketitle
\begin{abstract}
Physical mechanisms that can influence rotation curves of spiral galaxies
are discussed. For dark matter studies, possible contributions due to magnetic
fields and non-Newtonian gravitational accelerations should be carefully acconted
for. We point out that magnetic fields are particularly important in outermost
parts of the disk. 
In the framework of general relativity the physical 
reason 
of an enhanced gravity in spiral galaxies depends on the assumed metric.
The additional gravity is provided  for Schwarzschild metric by  nonluminous 
mass, whereas
for Vaidya metric [1] by emission of radiative energy. In the latter case the
non-Newtonian acceleration displays $1/r$ behaviour. Also matter flows  
contribute to non-Newtonian gravity.

\end{abstract}
\PACS{98.52.Nr.,98.62.Dm.}
  
\section{Introduction}
The best evidence of enhanced gravity in galaxies 
is provided by
flat rotation curves of spiral galaxies, which do not decay in a Keplerian
way even far from the rotation axis. From simple Newtonian formula for centripetal
acceleration,
\begin{equation}
{v_{rot}^2 \over r} = {G M(r) \over r^2}, \end{equation}
one finds that the total mass within radius $r$ grows with r as $M(r) \sim r v_{rot}^2$.
The linear growth of mass is customerly attributed to an invisible component, 
referred to as dark matter.

The problem of dark matter has its beginning in the observational determination 
by Zwicky of
dynamical mass of the Coma cluster of galaxies. The gravitational mass inferred
by Zwicky from the motion of individual galaxies in the cluster exceeded by
a factor of a few hundred the mass obtained by measuring luminosities assuming typical
value of mass to light ratio. 
Later a discrepancy between dynamical and luminous mass has been found
in spiral galaxies and galaxy clusters. 

Our aim here is to point out that the conclusion as to the existence of dark
matter inferred from rotation curves is not inescapable, but based on some 
unspelled assumptions. It holds in Newtonian 
gravity provided any role of magnetic fields is negligible.
In the framework
of Einstein's gravity for it to hold one {\it implicite} assumes space-time geometry to
be given by the Schwarzschild metric. 
It is often assumed that galactic gravitational field, as very weak
one, is adequately described by Newtonian gravity. We will show, employing Vaidya
metric, that the inverse problem, of reconstructing galactic gravity
given rotation velocity, has also other solutions. One encounters here ambiguity, 
which can only be resolved by physical input. 

Recent observations of dearth of dark matter in elliptical galaxies [2] suggest
that there may be
more unknowns involved in this problem.
One should also consider non-gravitational origin of the above discrepancy, 
namely due to 
magnetic fields in galaxies. The role of magnetic fields is likely very important 
in the outer disk region, where the galaxy rotation is detected by tracing hydrogen 
clouds. 

In order to firmly infer the amount of dark matter in spiral galaxies one should
subtract contributions to rotation curves generated by other forces (i.e.
magnetic fields) and
processes. It is certain that such contributions exist as in many galaxies
rotation curves show wiggly structure, as e.g. in our Galaxy. Such a structure
cannot be produced by WIMP gravity, as density of WIMPs is a monotonically
decreasing function of distance from the center of the distribution. Magnetic
influence and energy-flow-generated gravity can easily account for 
undulations of rotations curves. However, immediately arises a question how much
such forces/processes contribute to the bulk of rotation curves. Before gauging
this influence the inferred amount of dark matter in a galaxy is subject to 
substantial uncertainty. 

In the next section the problem of flat rotation curves of spiral galaxies
is briefly reminded. In sect.3 the role of magnetic fields is discussed. In
sect.4 gravity due to radiation flow is discussed with the use of Vaidya metric [1].
Finally in the last section we summarize important points once again.

\section{Enhanced gravity in galaxies}

One can formulate the problem of flat rotational curves precisely as follows: 
there is too much gravity compared to mass we can account for by counting stars and
measuring the amount of gas in galaxies. It is a nonrelativistic
custom  to attribute this enhanced gravity to invisible and (almost) undetectable
matter. In a popular Cold Dark Matter model, invisible mass is due to hypothetical weakly 
interacting massive particles - WIMPs. 

In general relativity, which is
supposed to be the theory of gravity, not only mass is capable of generating
gravitational field, but also energy or radiation flows induce gravity.
Interpreting gravitational acceleration in Newtonian terms, 
${\bf a}=-GM(r)/r^2 {\hat{\bf r}}$, for spherical symmetry one {\it implicite} assumes
the Schwarzschild interior metric
\begin{equation}
ds^2=e^{\nu}dt^2-e^{\lambda}dr^2-r^2d\Omega^2, \end{equation} 
with $e^{-\lambda}=e^{\nu}=1-2M(r)/r$ for matter  with negligible pressure, described by a dust
equation of state. 
In the weak field limit, which is 
appropriate for galactic fields, one obtains then the Newtonian acceleration. 

Typical rotational velocities of spiral galaxies, in the flat regime, are of
order of 100 km/s. One can thus infer the mass within radius $r$ to be
$M(r)=2.32\times 10^9 (v_c/100 km/s)^2r/kpc M_{\odot}$. For  Milky Way galaxy this gives 
within $30 kpc$ the mass $M_{MW}=3.37\times10^{11} M_{\odot}$ for $v_c=220km/s$. This high value 
of mass is thought to show us that the main component of mass
in our Galaxy, and in other galaxies, is nonluminous. Astronomers tried hard
to detect known nonluminous astrophysical objects that could form an invisible
population providing the missing mass. All attempts to account for it by dim  
stars, dead stars,  plasma or other forms of baryon matter have failed. 
The only viable candidate at the moment 
is particle dark matter composed of
WIMPs (or axions), forming an extended halo around galaxies. 
The radius of this halo is
presently unknown, but some observations suggest it is of order of 200 kpc.
Also, some cold gas in the form of molecular hydrogen, can contribute to dark matter, as it is 
very difficult to detect this component.

One can briefly summarize that the dark matter hypothesis is a Newtonian
solution of enhanced gravity problem in spiral galaxies with any influence of magnetic field
neglected.

\section{Magnetic fields and the rotation curves}

Magnetic fields are the most common phenomenon 
in spiral galaxies where we observe fields of regular and chaotic structure. 
The regular  structure has azimuthal and poloidal components. 
The poloidal component of the magnetic field is produced by the galactic 
dynamo effect.
The azimuthal component is induced from poloidal component by differential rotation. 
Regular fields created by such mechanisms may reach intensities of a few to 
several hundred microgauss. 

A question arises [3,4] whether these fields have any influence on the galactic 
rotation curves. 
In order to answer this question we must  investigate the Navier-Stokes 
equation with 
magnetic field,
\begin{equation}\rho[\frac{\partial \vec{v}}{\partial t}+(\vec{v}\nabla)\vec{v}]
= -\nabla p-\frac{M G \rho}{r^{2}}+\eta\Delta\vec{v}+(\xi+\frac{\eta}{3})
\nabla(\nabla \circ \vec{v})+\frac{1}{4 \pi}((\nabla\times \vec{B})\times
\vec{B}). \end{equation}
 We assume here the gravitational field of point mass located at 
the center of galaxy with $M\sim 2 \cdot 10^{44}g$, which is a good 
approximation for our exploratory calculation. In stationary galactic disk
we can neglect radial velocities and viscosity and we will compare gravitational
and magnetic field forces.
Rough estimate shows that for gas clouds of density $\rho \sim 
10^{-25}g/cm^{3}$ at the radius of a few tens kpc, $r\sim 3 \cdot 10^{22}cm$, from
the galactic 
center, and for magnetic fields of a few $\mu G$, $B\sim  10^{-5}G$, magnetic 
forces are comparable with gravitational forces. Gravitational accleration and
accleration due to magnetic effects are, respectively, 
\begin{equation}\frac{M \cdot G}{r^{2}}\sim 10^{-8}cm s^{-2},\end{equation}
\begin{equation}\frac{B^{2}}{\rho\cdot r}\sim 3 \cdot 10^{-8}cm s^{-2}.\end{equation}

For magnetic effects to occur the
gas must be partially ionized. We know that at least a few percent of 
the hydrogen 
in galaxies is ionized. Therefore we may  expect, that the magnetic fields' 
influence on rotational curves is not negligible. 

The above order-of-magnitude
estimate shows that the magnetic influence is particularly important in the
outermost regions of the galactic disk, where the density of hydrogen is the
lowest. From eq.(3.3) we find that when density decreases by a factor of 100,
magnetic fields on the scale of $1 \mu G$ can overwhelm gravity! Let us remind
that most of the dark matter contribution to galaxy rotation comes from the
outskirts. Any unaccounted for magnetic field contribution can completely 
corrupt dark mass measurement. One should also keep in mind that magnetic fields
on $\mu G$ order are expected to be ubiquitous in the intergalactic space in
clusters of galaxies.   

Taking only azimuthal component of magnetic field into account and assuming, 
that it depends only on its radial coordinate we can get a simple analytical 
form of this 
component which will flatten the rotation curve:
\begin{equation}\frac{(v_{\phi })^{2}}{r}=\frac{M G}{r^{2}}+
\frac{1}{4\pi\rho r}B_{\phi }\frac{\partial }{\partial r}(r B_{\phi }),\end{equation}
where $v_{\phi }=v_c= const$ is the rotational velocity of a galaxy.
The solution of this equation reads:
\begin{equation}B_{\phi }=+/-\frac{2 \sqrt{2 \rho \pi}\sqrt{0.5 (v_{\phi
})^{2}r^{2}-M G r-C}  }{r},\end{equation}
where $C$ is the integration constant.

The real  magnetic fields in spiral galaxies have all the components, poloidal and azimuthal, nonvanishing. 
Therefore the radial part of Navier-Stokes equation has the following form:
\begin{equation}\frac{(v_{\phi })^{2}}{r}=\frac{M G}{r^{2}}-\frac{1}{4\pi\rho
}\Bigg((\frac{\partial B_{r}}{\partial z}-\frac{\partial B_{z}}{\partial
r})B_{z}-B_{\phi }(\frac{1}{r}\frac{\partial }{\partial r}(r B_{\phi
})-\frac{1}{r}\frac{\partial B_{r}}{\partial \phi })\Bigg).\end{equation}
To assess a possible influence of such magnetic fields on rotation curves one should take into 
account the whole 3D structure of magnetic field.  

\begin{quote} \end{quote}

\begin{quote} \end{quote}
\begin{figure}[h]
\centering
\includegraphics[angle=0,height=0.4\textheight,width=0.8\textwidth]
{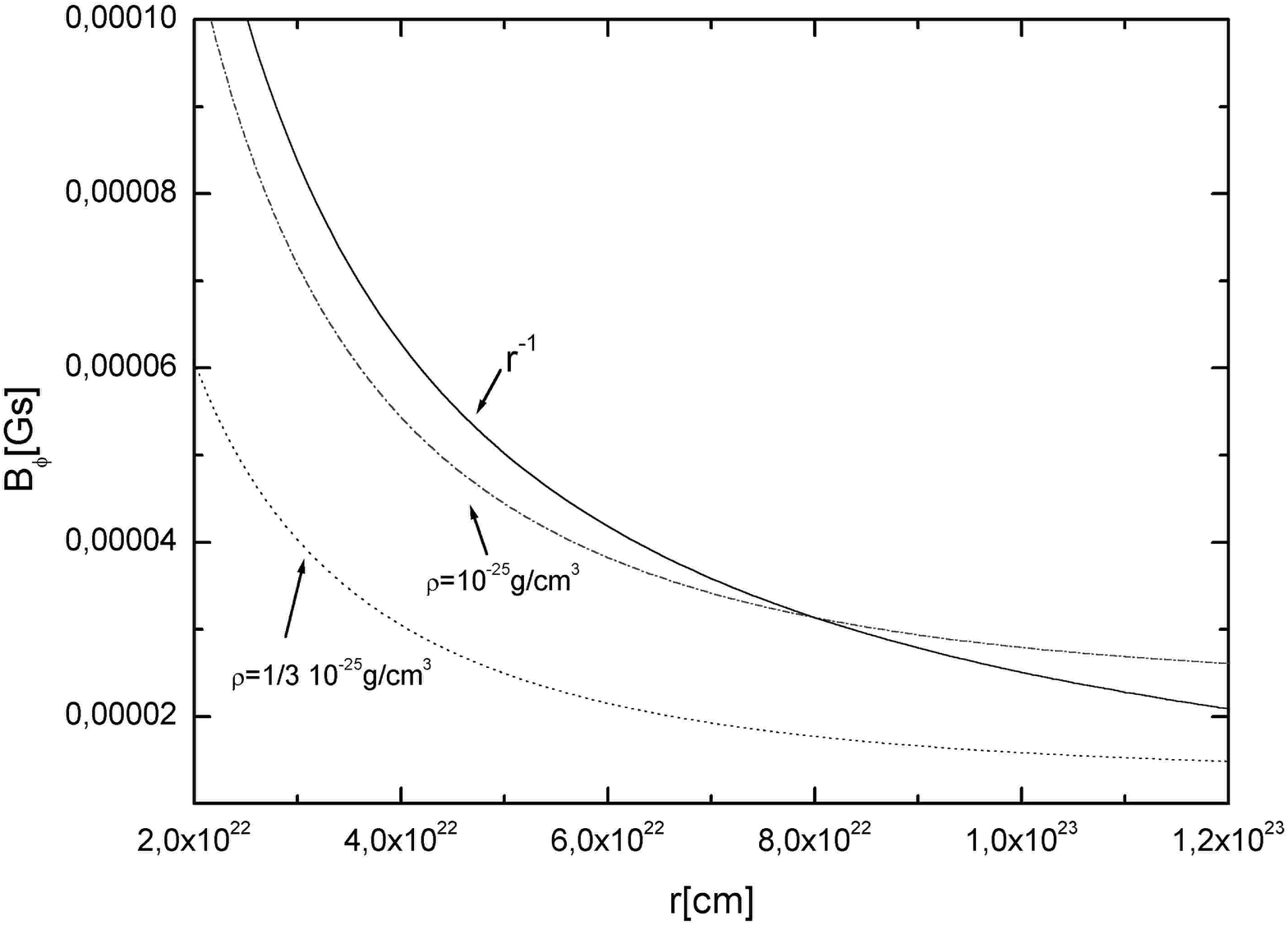}
\vspace{-0.04\textheight} 
\caption{The dashed-dotted line is a solution of equation (3.8) 
with $M=2\cdot 10^{44}g$, $v_{\phi }=220\cdot km/s$, 
$\rho =10^{-25}g/cm^3$,dotted line is a solution corresponding to density 
$\rho =0.333\cdot 10^{-25}g/cm^3$. Solid curve is a function  $B_{\phi} \sim 1/r$} \label{}
\end{figure}

\section {Gravity generated by radiation flow: the Vaidya metric}

The Schwarzschild metric describes strictly speaking gravity of cold spherically symmetric  
astrophysical body, 
such as a planet, dead star or a black hole. 
Gravity of  radiating
objects such as normal stars is only approximately described by the Schwarzschild metric.
There exists, however, exact solution of Einstein's field equations corresponding essentially
to a real star which emits radial flux of radiation, found by Vaidya [1]. In its original form, the
Vaidya metric is
\begin{equation}
 ds^2=({{\dot m} \over m'})^2(1-{2m \over r})^{-1}dt^2-(1-{2m \over
r})^{-1}dr^2-r^2d\Omega^2.\end{equation}
It corresponds to the space-time region outside the star, $r>r_0$, where $r_0$ is  
the stellar radius, and $m \equiv m(r,t)$.
This metric can be cast is a very elegant form employing the retarded time variable $u=t-r$, 
\begin{equation}
 ds^2=-(1-{2m \over r})du^2-2dudr+r^2d\Omega^2,\end{equation}  
as shown by Vaidya in Ref.[1].

The energy tensor corresponding to the Vaidya metric has non-zero $T^1_0$ component, which
describes the energy outflow carried away by massless fields. Let us consider the energy 
tensor for directed flow of radiation [5] in the form
\begin{equation}
 T_{\mu}^{\nu}=\rho v_{\mu}v^{\nu},\end{equation}
where $\rho $ is the energy density of radiation, and the fourvector $v^{\mu}$ is null,
$v^{\mu}v_{\mu}=0$. For the radial outflow, $v^2=v^3=0$, and $T_2^2=T_3^3=0$.
The metric (4.1) is a particular example of a general non-static spherically symmetric metric
[5],
\begin{equation}
ds^2=e^{\nu(r,t)}dt^2-e^{\lambda(r,t)}dr^2-r^2d\Omega^2. \end{equation}
The Einstein's field equations are [5]:

\begin{equation}-8\pi T_0^0=-{1 \over r^2}+e^{-\lambda}({1 \over r^2}-{\lambda'
\over r}),\end{equation}
\begin{equation}-8\pi T_1^1=-{1 \over r^2}+e^{-\lambda}({1 \over r^2}+{\nu'
\over r}),\end{equation}
\begin{equation}
-8\pi T_2^2=-{1 \over 4} e^{-\nu} (2{\ddot \lambda}+{\dot \lambda}({\dot \lambda}-
{\dot \nu }))
+{1 \over 4}e^{-\lambda}(2\nu''+\nu'^2-\lambda'\nu'+2{\nu'-\lambda' \over
r})=-8\pi T_3^3,\end{equation}
\begin{equation}
-8\pi T_{01}=-{1 \over r}{\dot \lambda}.\end{equation}

Let us introduce the mass function $m(r,t)$ through $e^{-\lambda(r,t)}=1-2m(r,t)/r$. 
From the null
condition, $v_{\mu}v^{\mu}=0=-e^{\lambda}(v^1)^2+e^{\nu}(v^0)^2$ we find
$e^{(\nu-\lambda)/2}T_1^0+T_0^0=0$ which gives
\begin{equation}
 e^{-\lambda'/2} m'+e^{-\nu /2}{\dot m}=0.\end{equation}
This allows us to express the function $e^{\nu}$ through $m(r,t)$ and its derivatives,
and to
write the metric (4.1) in the form (4.2) given by Vaidya.

The physical interpretation of Vaidya's metric is straighforward. In the weak field 
limit we find ${\dot m}+m'=0$ and the energy flux flowing out of a sphere of radius r is
\begin{equation}
 T^{01}=-{m'^2 \over 4\pi r^2 {\dot m}}=-{{\dot m} \over 4\pi r^2}. \end{equation}
Hence $m(r,t)$ is the radiation energy inside this sphere,
\begin{equation}
m(r,t)=\int 4\pi r^2T_0^0dr.\end{equation}
The function ${\dot m}$ is the rate of energy emission, or total luminosity, and 
$m'=4\pi r^2T_0^0$. We should also include the radiation source, located at the origin, 
which loses energy at a rate ${\dot M}(0,t-r)={\dot m}(r,t)$.

The most important result is the gravitational acceleration in the weak field limit, 
$e^{\nu}\approx 1+\nu$.
From general expression
we find for the  metric (4.4) the Einstein's formula
\begin{equation}
{\ddot r}=-{1 \over 2}\nu'. \end{equation}
When applied to the Schwarzschild metric in the weak field limit,
with $\nu = -2M/r$, it gives the Newtonian acceleration.

For the Vaidya metric from eq.(4.6) we have 

\begin{equation} \nu'={2 \over r}[(e^{\lambda}-1)+e^{\lambda}(m'-{m \over r})].\end{equation}
In the weak field limit we find
\begin{equation}{\ddot r}= -{m' \over r}-{m \over r^2}.\end{equation}
This expression shows that there appears a non-Newtonian acceleration
\begin{equation}
a_L=-{m' \over r}=G{{\dot m} \over rc}, \end{equation}
which is inversely proportional to the distance. Far from the center, $a_L$ becomes
dominating. For radiating body, with energy flowing out of the central mass, ${\dot m}<0$ 
and the the acceleration (4.14) produces an attractive force which becomes stronger that usual
Newtonian gravitation. 

The additional gravitational attraction due to radiation emission
implied by the Vaidya metric was first discussed by Lindquist et al.[6]. It gives an 
explicit example of non-Newtonian gravitational force resulting from Einstein's gravity
theory for a realistic metric in the weak field limit. Thus the notion that Newtonian 
acceleration
is the only  weak-field limit of general relativity is inaccurate. 

The formula (4.14) can be generalized to the case of galactic wind which is a
radial matter outflow,
 \begin{equation}
a_{wind}=G{{\dot m} v_{wind} \over rc^2}. \end{equation}
Here $v_{wind}$ is the radial velocity of the wind and ${\dot m}<0$ is the mass
loss rate due to wind. Please note that the above formula (4.15) is also valid
for radial accretion, with radial infall velocity $v_r<0$. Since for accretion
the mass increases, ${\dot m}>0$, the induced acceleration is also directed
inward, as for the wind. One can thus conclude that radially oscillating shell
of gas would always produce gravitational attraction, both in expansion and
contraction phase.

\section{Discussion}

 Astronomers tend to
consider the Newtonian solution of the enhanced gravity problem in spiral
galaxies to be the only one compatible with general
relativity. One can encounter statements that any non-Newtonian gravitational
acceleration in galaxies, as e.g. employed by Milgrom in his model of galaxy
gravity [7], would necessarily require modifications of Einstein's gravity theory. 
We have given here an example that the statement 
that non-Newtonian gravitational acceleration,
$a \sim 1/r$, is incompatible with
general relativity, is not true. 
The acceleration (4.14) can be shown to produce flat rotational curves of spiral
galaxies. The centripetal acceleration when the non-Newtonian acceleration dominates, is
\begin{equation}
{v_{rot}^2 \over r} = {G {\dot m} \over cr}, \end{equation}
which allows us to calculate the source luminosity $L=-{\dot m}$. For $v_{rot}=100 km/s$
$L\sim 10^{52} erg/s$. Hence the problem of enhanced gravity in spiral galaxies 
with the Vaidya metric changes to the problem of the energy source and the physical nature 
of its emission. Physically, it is very different from the Newtonian solution, which is
the nonluminous matter.

Presently it is a standard assumption that galactic dynamics is governed by dark matter.
To prove the dark matter hypothesis a number of experiments start to search for neutralino, the
best supersymmetry candidate for WIMP. Also, astrophysical observations of dark matter in
elliptical galaxies have been attempted by PN.S collaboration [2] with planetary nebulae as
a tracer of gravity. Surprisingly, gravity of those galaxies is adequately described by
luminous matter only , a result described as a "missing missing mass" problem [2]. If the
results obtained by PN.S collaboration are correct than dark matter in elliptical galaxies
is at least differently distributed than in spiral galaxies, with only trace amount inside
inner 5-6 effective radii. 

A radical proposal is the Milgrom's modified Newtonian dynamics (MOND) hypothesis, 
which postulates new gravitation law
for very weak accelerations. This proposal, employed as a phenomenological model, is quite
successful in explaining spiral galaxies dynamics. MOND also explains the dearth of dark
matter in elliptical galaxies [8] observed by Romanowsky et al. [2]. The radial dependence
of the MOND acceleration is the same as in non-Newtonian acceleration $a_L$
(4.14) for the Vaidya metric.

One can notice that the Vaidya metric is an example of metric 
considered recently by Lake [9] that can give flat rotation curves of spiral
galaxies.

We  have shown here that magnetic fields in spiral galaxies can play a crucial role in 
determining the rotation curves. The solution (3.6) shows that magnetic field which fully accounts
for a flat rotation curve has toroidal component compatible with observed magnetic fields in
spiral galaxies. It may not be a good approximation to completely suppress magnetic field 
influence when studying the physical origin of flat rotation curves. In realistic description
observed magnetic field influence should be subracted before fitting gravitational potential
generated by assumed dark matter halo.
The importance of magnetic field contribution to flat rotational curves of
spiral galaxies has been recently discussed in Ref.[10].

It is also worth to notice that pure 
magnetic mechanism of flat rotation curves in spiral galaxies could explain
simultaneousely  why 
these curves 
in elliptical galaxies are Keplerian. It is because there are no regular 
magnetic fields in elliptical galaxies.

\end{document}